\documentclass[a4paper,11pt]{article}
\usepackage{pos}
\usepackage{subcaption}
\usepackage{wrapfig}
\title{Development of a Multi-Purpose Optical TPC for Neutron-Induced Reaction Studies at SARAF}
\ShortTitle{Optical TPC for neutron-induced reactions at SARAF}

\author*[a]{R. Felkai}
\author[a]{M. Borysova}
\author[a]{L. Moleri}
\author[a]{J. Pienaar}
\author[a]{D. Vartsky}
\author[a]{A. Breskin}
\author[b]{I. Mor}
\author[b]{L. Weissman}
\author[a]{S. Bressler}

\affiliation[a]{Department of Particle Physics and Astrophysics, Weizmann Institute of Science,\\
Herzl St 234, Rehovot, Israel}

\affiliation[b]{Soreq Nuclear Research Center,\\
Yavne, Israel}

\emailAdd{ryan.felkai@weizmann.ac.il}

\abstract{Neutron-induced reactions play a central role in stellar and Big Bang nucleosynthesis models.
Yet many of the cross sections remain poorly constrained at astrophysically relevant energies.
To address these needs, we are developing a multi-purpose Optical Time Projection Chamber (OTPC) optimized for precision neutron-reaction studies at the Soreq Applied Research Accelerator Facility (SARAF) upcoming high-intensity, time-of-flight neutron beam.
The detector combines a drift chamber filled with CF$_4$-based scintillating gas mixtures, fast photodetectors for prompt scintillation detection, and high-speed optical readout of avalanche-induced secondary scintillation to enable full 3D reconstruction of charged-particle tracks.
A prototype system has been assembled and tested.
This has enabled systematic characterization of electron drift velocity, charge and light amplification, and 2D optical imaging of alpha-particle tracks in an Ar/CF$_4$ gas mixture.
These studies guide the design of a larger, fully integrated OTPC system intended for operation at SARAF.
In parallel, we are exploring advanced image sensors to further enhance tracking resolution.
We report on recent progress with the prototype and outline the next steps toward commissioning the full system.}

\FullConference{19th International Conference on Topics in Astroparticle and Underground Physics (TAUP2025)\\
24–30 Aug 2025\\
Xichang, China\\}

\begin{document}
\maketitle

\section{Introduction and motivations}

\begin{wrapfigure}{r}{0.48\textwidth}
  \centering
  \includegraphics[width=\linewidth]{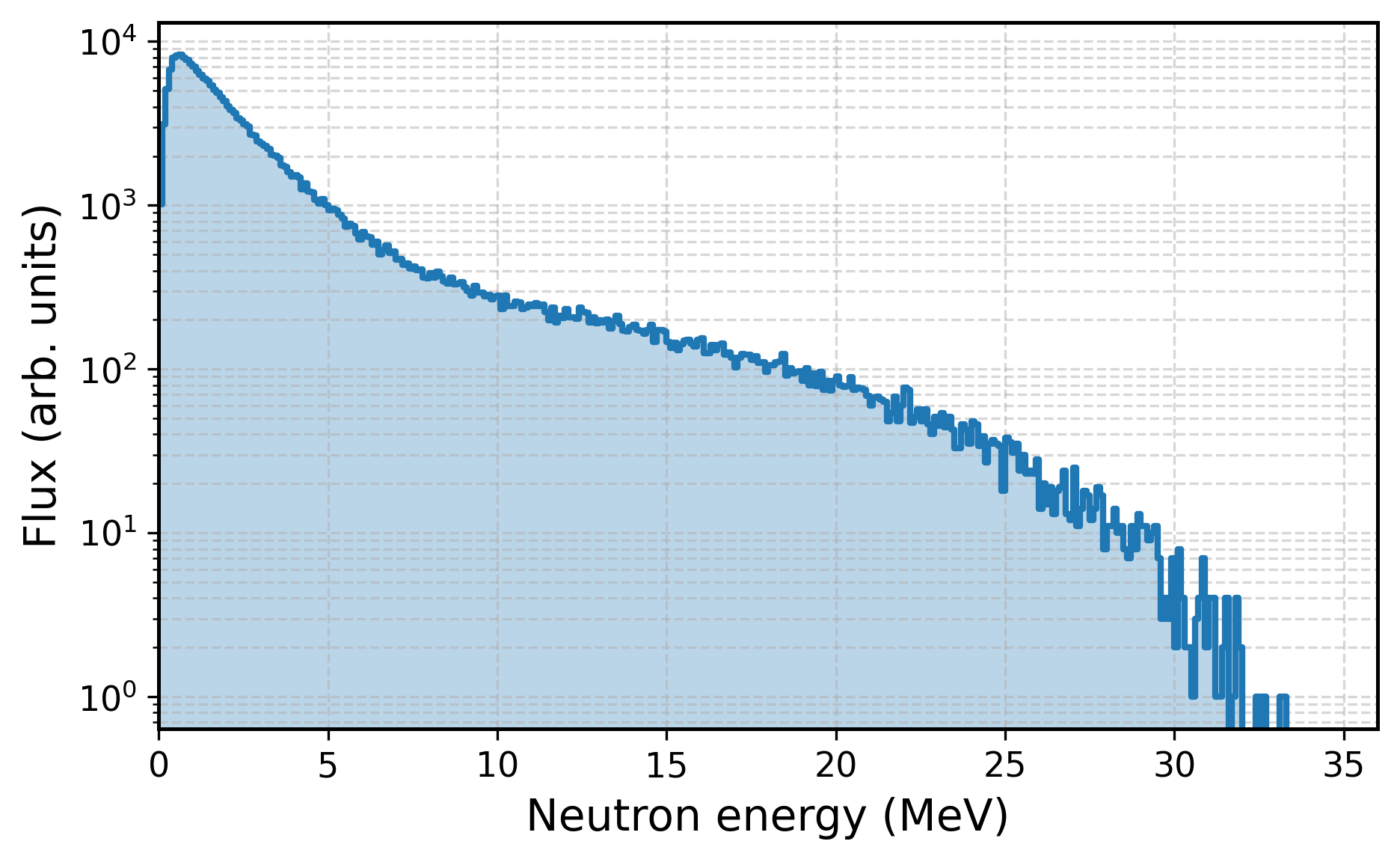}
  \caption{Expected neutron energy spectrum of the beam at SARAF-II, obtained for protons on a liquid gallium-indium jet target (unpublished).}
  \label{fig:ebeam}
\end{wrapfigure}
Optical Time Projection Chambers (OTPCs)~\cite{Gai2020} are versatile gaseous detectors that provide 3D tracking and nearly full solid-angle acceptance.
These capabilities make them a popular choice for many experiments, e.g. rare-event searches in underground laboratories.

We are developing an OTPC as a general-purpose detector for neutron-induced reaction studies at the Soreq Applied Research Accelerator Facility (SARAF) \cite{Mardor2018}.
The upcoming SARAF-II upgrade will deliver a broad energy, intense pulsed fast-neutron beamline (see Fig.~\ref{fig:ebeam}).
Time-of-flight (TOF) will allow neutron energy selection from $\sim$10~keV to tens of MeV.
In this environment, an OTPC offers event-by-event kinematics and particle identification for all charged products in a single device, which is attractive for cross-section measurements and for background rejection.
Our science scope includes neutron-induced fission followed by lighter reactions of astrophysical interest (such as $^{7}\mathrm{Be}(n,\alpha)\alpha$, $p(n,\gamma)d$, and $^{16}\mathrm{O}(n,\alpha)^{13}\mathrm{C}$).

Although our measurements will be performed at an accelerator, the detector technology and related challenges overlap with those of TPCs for most applications.

\section{Detector concept and prototype}

\begin{figure}[b]
  \centering
  \begin{subfigure}[t]{0.39\linewidth}
    \centering
    \includegraphics[height=0.2\textheight]{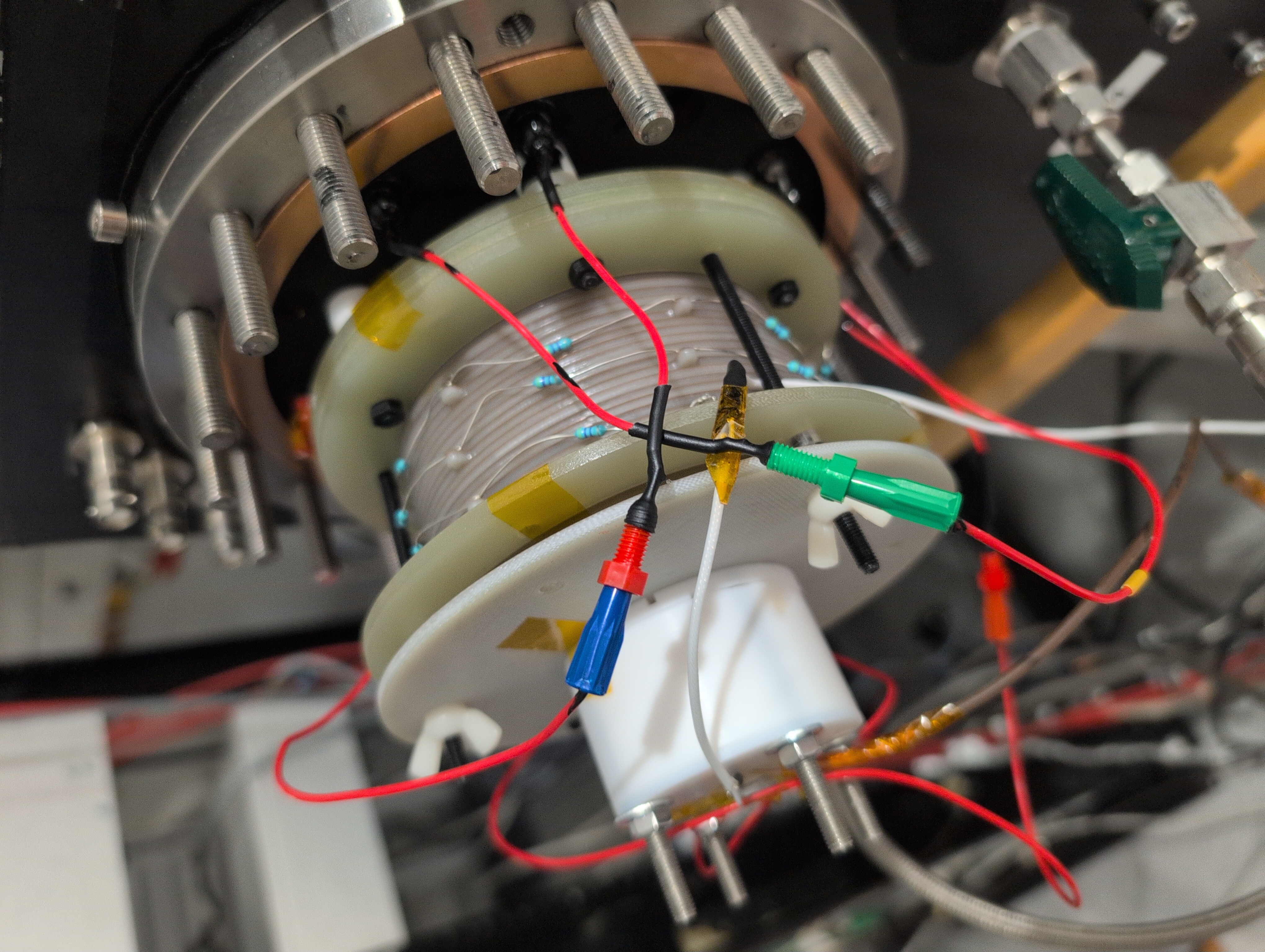}
  \end{subfigure}
  \begin{subfigure}[t]{0.59\linewidth}
  	\raggedleft
    \includegraphics[height=0.2\textheight]{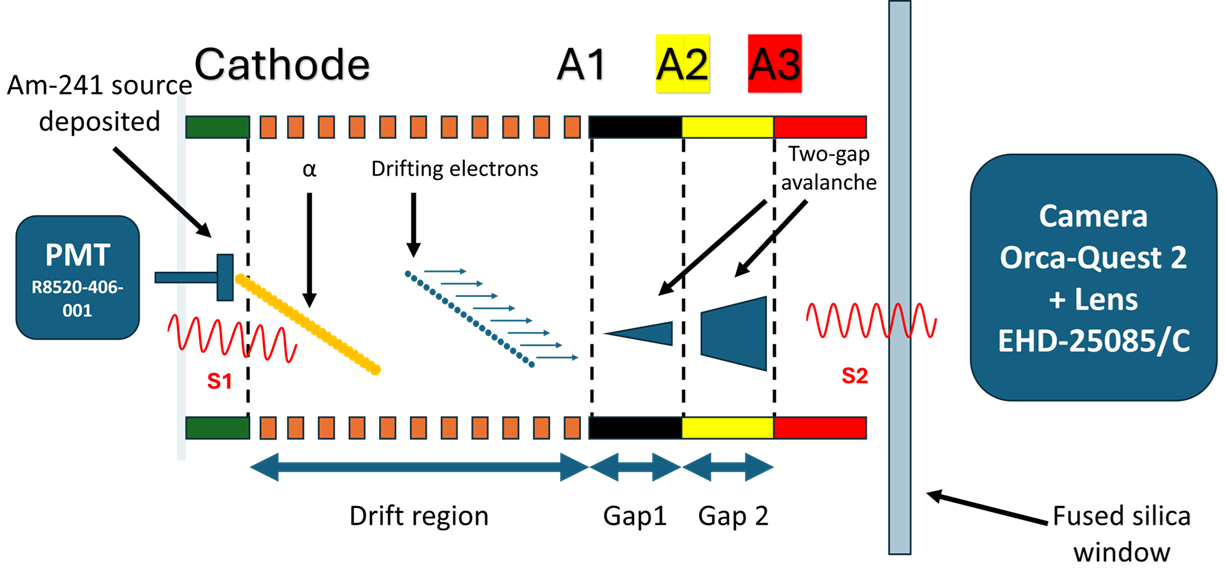}
  \end{subfigure}
  \caption{(left) Picture of the flange-mounted TPC assembly, viewed from the cathode side; (right) schematic of the assembly.}
  \label{fig:detector}
\end{figure}

We assembled a compact OTPC at the Weizmann Institute to validate all subsystems, including charge drift, charge multiplication and light production, light and charge readout, and 2D track imaging.
A picture of the detector assembly and the corresponding schematic are shown in Fig.~\ref{fig:detector}.

Ar/CF$_4$ (95:5) at 1~bar is used as the baseline gas.
This mixture offers high electroluminescence yield, mostly in near-UV and visible/red with components extending into the near-IR, well matched to modern CMOS sensors~\cite{Amedo2023}.
It also provides moderate operating voltages to achieve stable operation at 1~bar.

The detector active volume is defined by a cylindrical field cage (10~cm diameter and $\sim$5~cm drift length, leading to an active volume of 0.4~L) consisting of 21 FR4 rings clad on one side with a thin copper layer.
This geometry is sufficient to contain $\sim$5~MeV $\alpha$ tracks at 1~bar in Ar/CF$_4$ (95:5).
The electric field in the drift region is determined by two circular mesh-electrodes at the ends of the field cage.
These are made of woven stainless-steel wires stretched on FR4 frames (500~$\mu\mathrm{m}$ pitch and 50~$\mu\mathrm{m}$ wire diameter, leading to an optical transparency of 81\%).

Charge multiplication and light production are provided by a two‑step parallel-mesh multiplier~\cite{Breskin1988} made of three similar steel meshes (marked as A1, A2, A3 in Fig.~\ref{fig:detector}).
A1-A2 and A2-A3 define two amplification gaps, each of which can be configured with either 1.6~mm or 3~mm spacing.

A low‑noise charge‑sensitive preamplifier (Naicam CB200PC or Cremat CR-11X, depending on the desired gain) is connected to the mesh on which the electrons are collected: A1 in drift-only, A2 in a single-gap configuration, and A3 in the nominal two-gap configuration.
An in‑vessel PMT (Hamamatsu R8520-406-001) records both the prompt (S1) and avalanche (S2) scintillation photons generated by the primary energy deposition of the $\alpha$ in the gas and by the charge multiplication, respectively.
The signals from the different sensors were digitized by an oscilloscope, recorded and processed offline.
For track imaging we use a quantitative CMOS camera (ORCA-Quest 2) coupled to an ultra-fast photographic lens (f/0.85).
The lens is focused on the amplification plane so the recorded image is a 2D projection of the avalanche light from the two gaps.
The camera synchronization with the DAQ has not yet been streamlined.

The studies were conducted with a spectroscopic $^{241}\mathrm{Am}$ alpha source.
The source was located right behind the cathode, emitting alphas over $2\pi$ with an activity of $\sim$30~Bq, compatible with the camera timing limitations.

\section{Preliminary Results}

\begin{wrapfigure}{R}{0.48\textwidth}
  \centering
  \includegraphics[width=\linewidth]{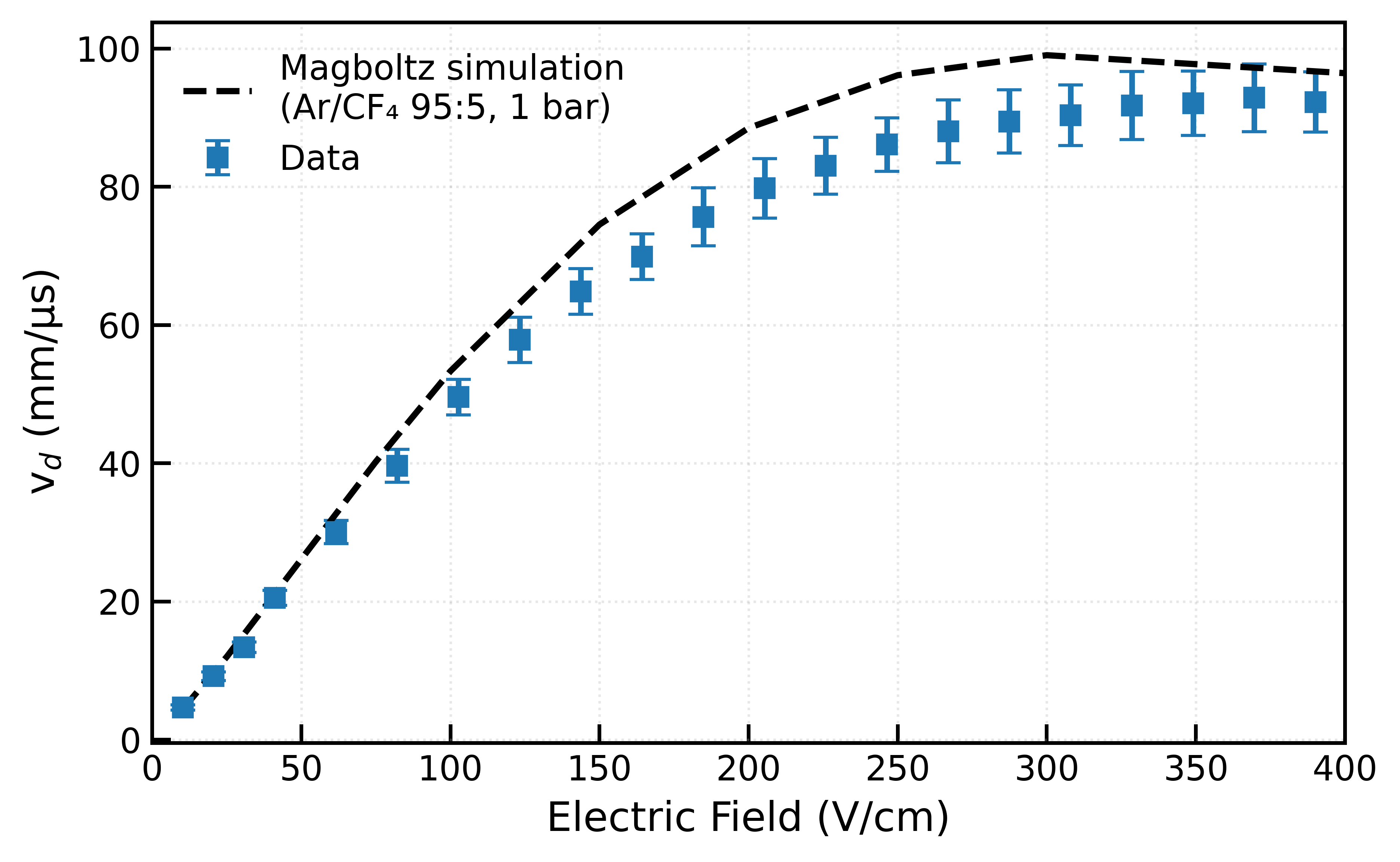}
  \caption{Electron drift velocity as a function of the drift field in Ar/CF$_4$ (95:5) at 1~bar. The dashed curve is a Magboltz simulation for the same mixture~\cite{Biagi1999}. The small discrepancy can be attributed to systematic uncertainties on the exact mixture and on the unmonitored impurity level.}
  \label{fig:driftvel_data}
\end{wrapfigure}
The drift velocity was measured with the detector in drift-only mode: the charge-sensitive preamplifier (CSP) was connected to the first anode (A1) with the field reversed on the amplification gaps to ensure electron collection on A1.
Because the alpha source was located behind the drift cathode mesh, each track starts at the cathode itself.
The drift of the free electrons induces a signal on the CSP at A1.
The signal rises until all electrons have reached A1, so the signal duration gives a measure of the drift time of electrons and thus the drift velocity (for the known drift length, see Fig.~\ref{fig:detector}).
A scan of the electron drift velocity as a function of the drift field for our gas mixture is shown in Fig.~\ref{fig:driftvel_data}.
For a typical drift field of 200~V/cm, an electron drift velocity of 80~mm/$\mu\mathrm{s}$ was measured. 

The absolute isotropic photon yield is obtained on an event-by-event basis.
First, the baseline-restored and zero-suppressed PMT waveform is integrated and converted to a number of photoelectrons using the single-photon calibration value; then the corresponding photon yield is calculated using the reported emission spectrum for Ar/CF$_4$~\cite{Brunbauer2018,Brunbauer2025}, the PMT quantum efficiency (provided by the manufacturer) and a solid angle calculation.
The relative photon yield (per secondary ionization) is simply the ratio of the absolute yield to the charge obtained from the corresponding CSP waveform.
Possible light reflections are not included; it is estimated that their inclusion would marginally decrease the photon yield.

\begin{figure}[t]
  \centering
  \begin{subfigure}[t]{0.49\linewidth}
    \centering
    \includegraphics[width=\linewidth]{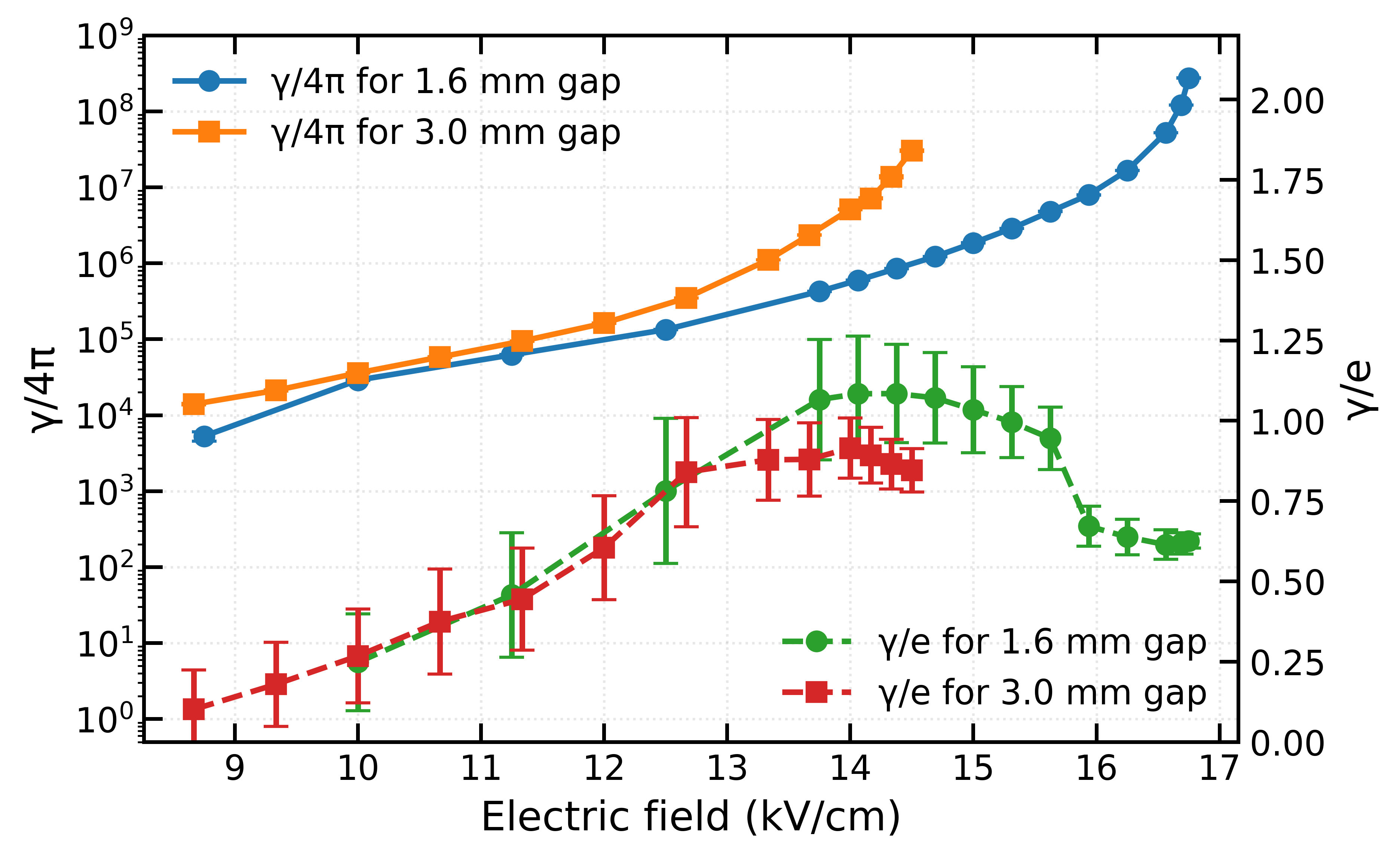}
  \end{subfigure}\hfill
  \begin{subfigure}[t]{0.49\linewidth}
    \centering
    \includegraphics[width=\linewidth]{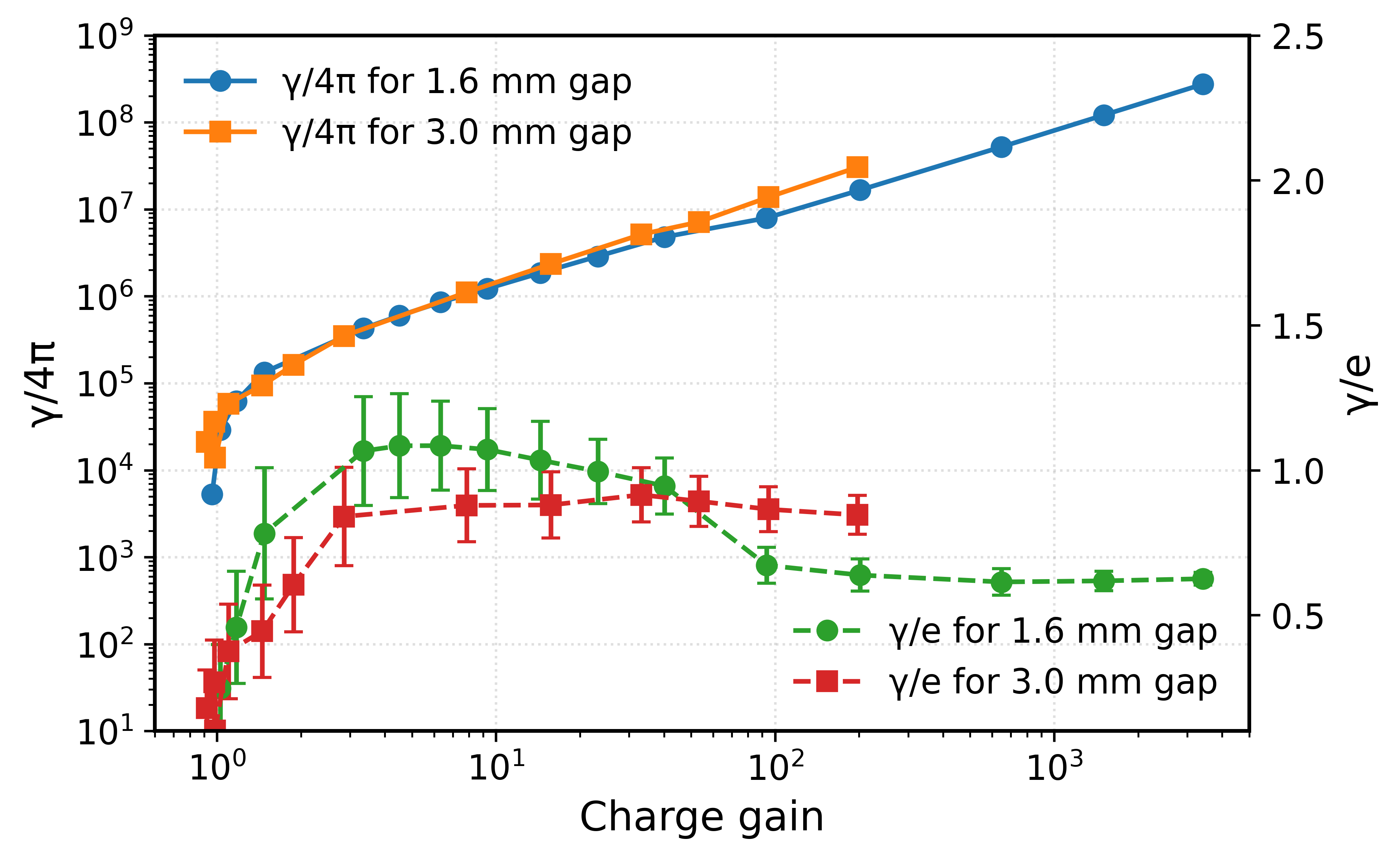}
  \end{subfigure}
  \caption{Characterization of the scintillation performance of the first gap for two thicknesses (1.6 and 3~mm). (left) Absolute and relative (per avalanche electron) photon yield as a function of the electric field. (right) Absolute and relative photon yield as a function of the charge gain, displayed on a logarithmic scale. Uncertainties on the absolute photon yield are omitted for readability.}
  \label{fig:perf_data}
\end{figure}

A single amplification gap was characterized for two thicknesses; the absolute and relative photon yields are shown in Fig.~\ref{fig:perf_data}.
At the highest fields, the absolute yield rises faster than exponentially while the relative yield drops, indicating that secondary effects (such as photon feedback) start driving electrical instabilities.

\begin{figure}[t]
  \centering
  \begin{subfigure}[t]{0.49\linewidth}
    \centering
    \includegraphics[width=\linewidth]{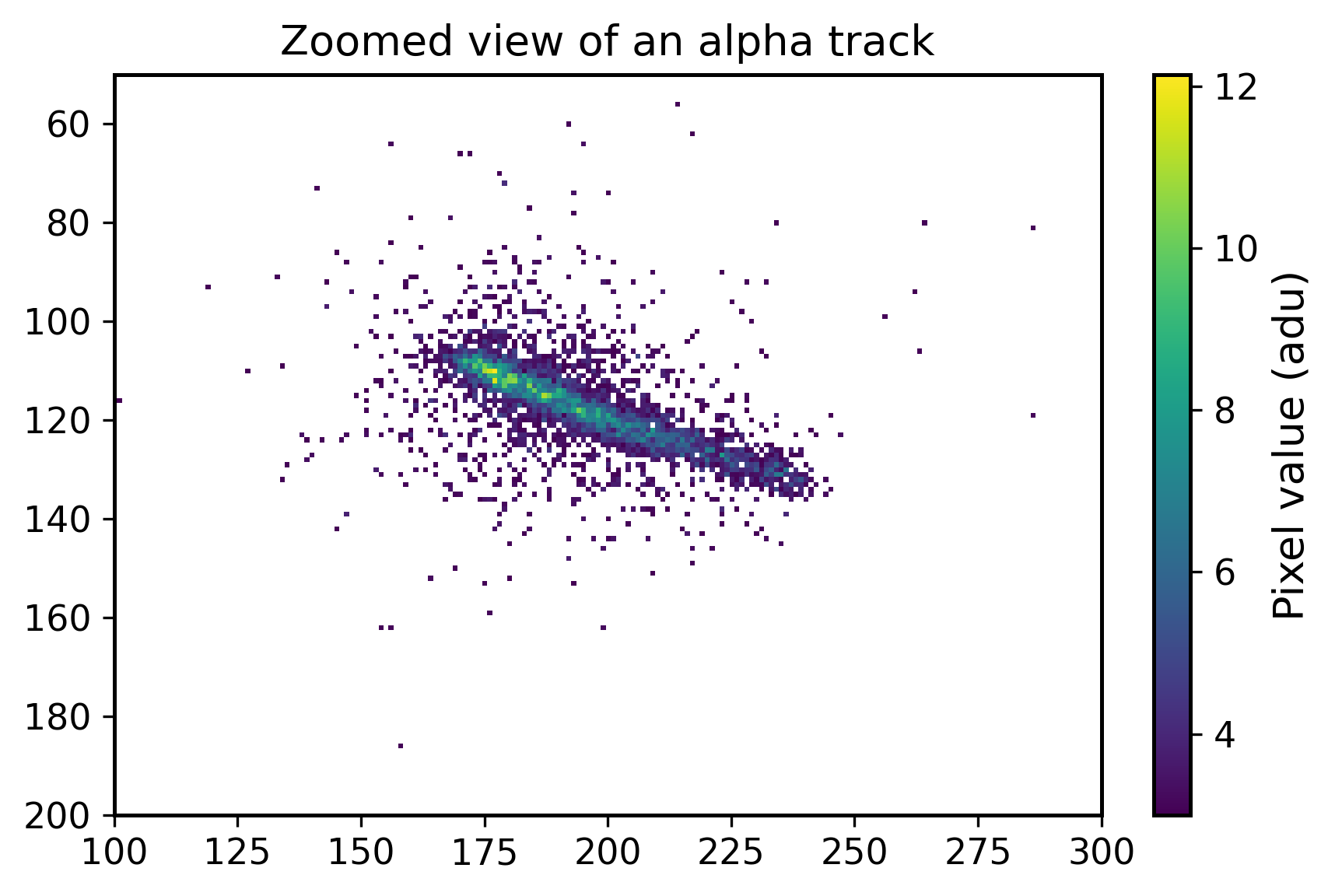}
  \end{subfigure}\hfill  
  \begin{subfigure}[t]{0.49\linewidth}
    \centering
    \includegraphics[width=\linewidth]{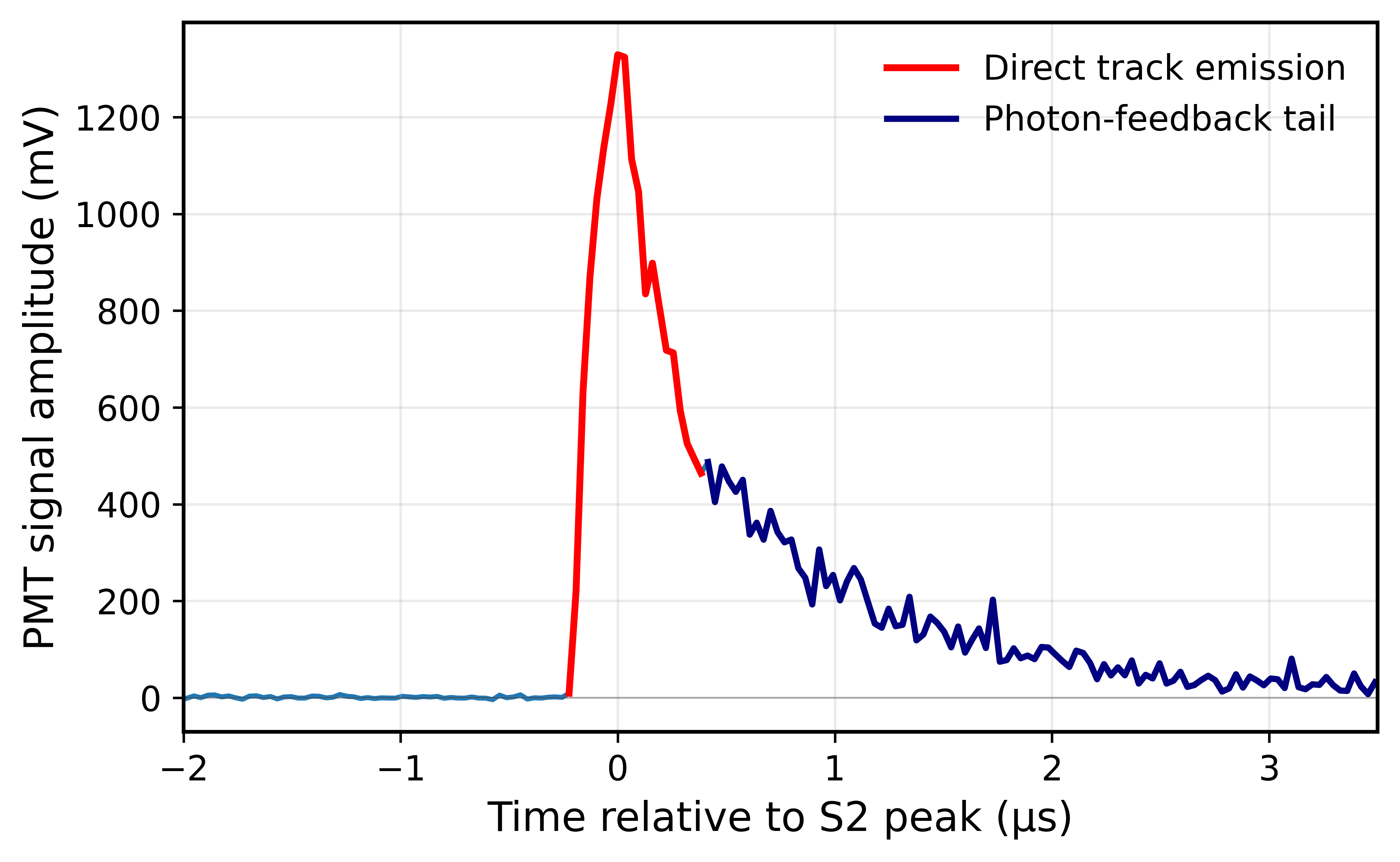}
  \end{subfigure}
  \caption{A single alpha track event. The left panel shows a zoom over a frame taken by the CMOS camera; the signal induced on the PMT by this track can be seen on the right panel; the part in red corresponds to the expected direct track emission according to the measured electron drift velocity, while the later dark-blue part is attributed to photon-feedback.}
  \label{fig:track}
\end{figure}

An early look at alpha tracks captured by the CMOS sensor was performed in a two-gap configuration.
This was a technical demonstration of synchronizing the camera with the PMT for offline analysis using the camera timing output.
It is planned to reconstruct tracks in 3D by combining the PMT signals and the camera frames as outlined in Ref.~\cite{Miernik2007}.
The left panel of Fig.~\ref{fig:track} shows a track (rebinned 8~$\times$~8 to enhance the signal-to-noise and taken with a 9~ms exposure) with the corresponding PMT waveform in the right panel.
The long tail is consistent with the aforementioned secondary effects; because the camera integration time largely exceeds the electroluminescence time profile, these effects contribute to the track luminosity.

\section{Next-generation detector and conclusion}

\begin{figure}[t]
  \centering
  \begin{subfigure}[t]{0.49\linewidth}
    \centering
    \includegraphics[height=0.25\textheight]{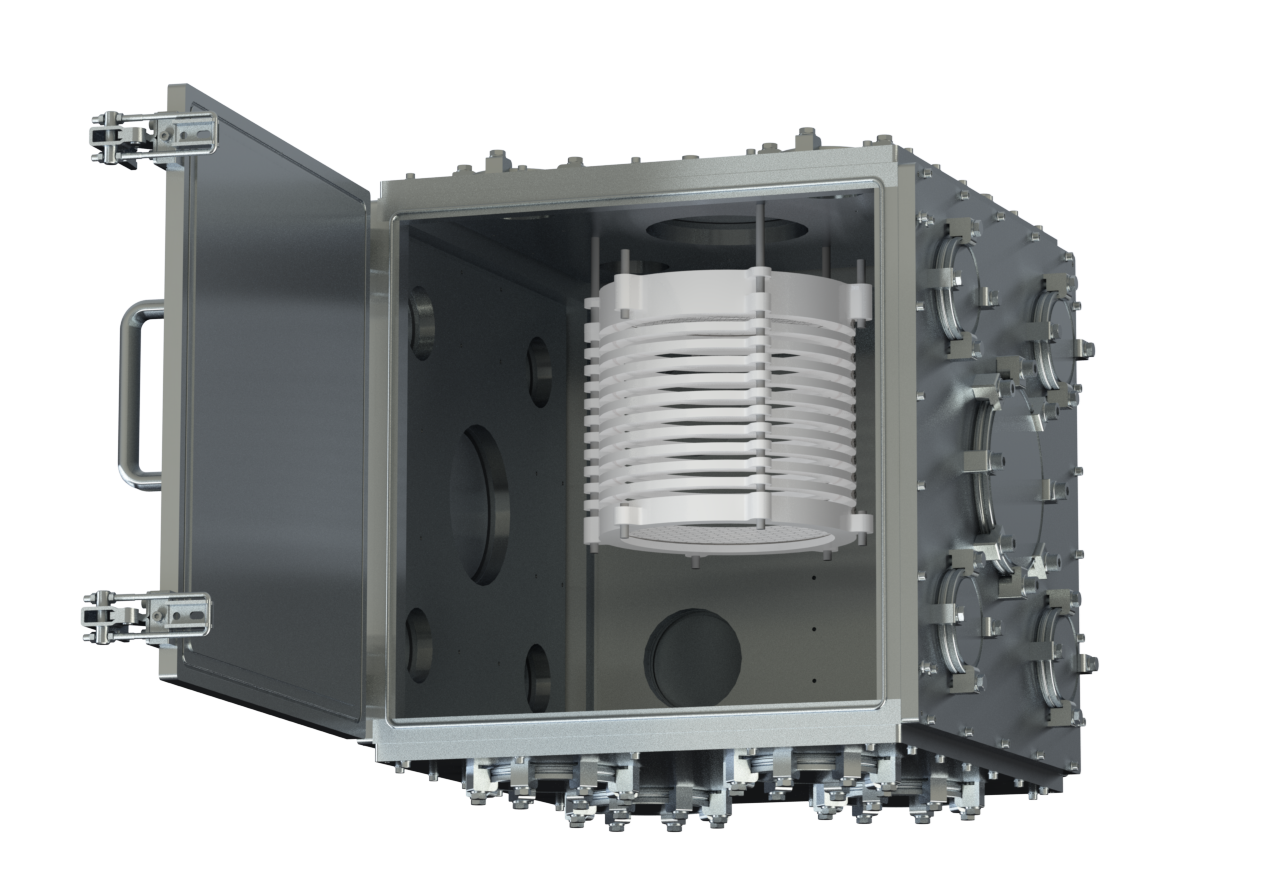}
  \end{subfigure}
  \begin{subfigure}[t]{0.49\linewidth}
  	\centering
    \includegraphics[height=0.25\textheight]{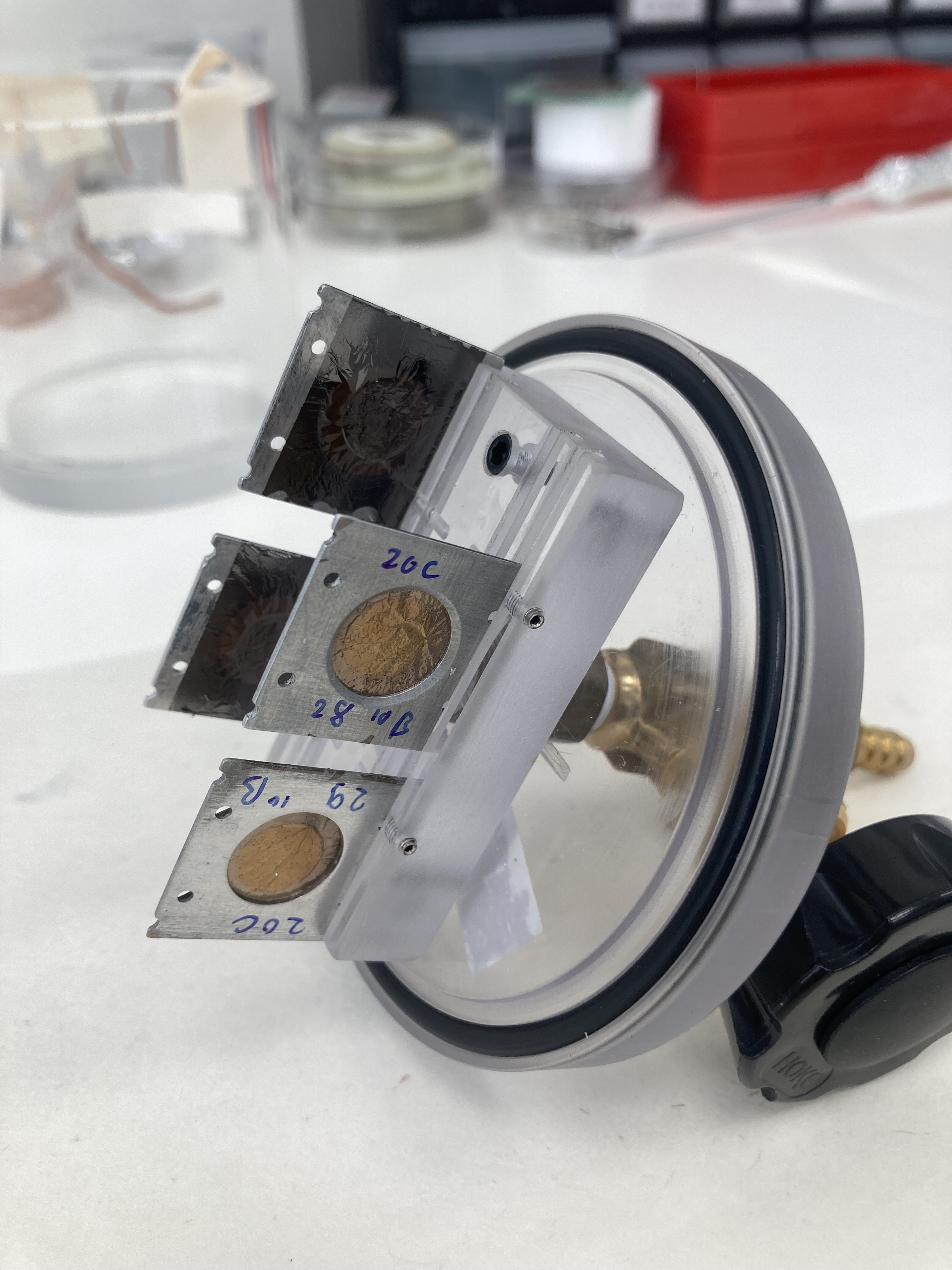}
  \end{subfigure}
  \caption{(left) CAD rendering of the full TPC assembly (field cage, cathode, and amplification stages) inside the custom vacuum vessel. (right) $^{10}$B targets on carbon backings have been produced for in-vessel measurements with a neutron source.}
  \label{fig:nextgen}
\end{figure}

One of our endeavours is the construction of a larger detector, incorporating the lessons learned from the current study and implementing modern design choices.
The core philosophy is to maximize flexibility and modularity with the goal of operating the detector in different modes: with neutron sources, near a nuclear reactor and, finally, at the SARAF-II beamline.
The detector will feature a larger drift length of about 20~cm, sized to contain $^{241}\mathrm{Am}$ alpha tracks at pressures down to 150~mbar. 
The field cage and amplification designs are finalized and in production while the custom-made vessel has already been produced; a CAD drawing of the full assembly can be seen on the left panel of Fig.~\ref{fig:nextgen}.
The dedicated gas system that will be paired with the detector will allow mixing of up to four gases, with the pressure tunable from 10~mbar to 1~bar.
We will study various mixtures of Ar/CF$_4$ and He/CF$_4$ while keeping the possibility of adding a small N$_2$ admixture.
It was shown~\cite{Margato2012, Brunbauer2018} that this small addition enhances the near-UV part of the spectrum at the expense of the visible band, which can improve track brightness for specific optical sensors.

The initial campaign for commissioning will be performed with $^{10}$B targets that have already been produced (Fig.~\ref{fig:nextgen}).
A simulation framework based on Geant4~\cite{Geant4-2003,Geant4-2016} is being developed to estimate the expected neutron-induced background in different configurations.
A fully time‑stamped 3D optical readout is planned in a follow‑up iteration using a Timepix3-based camera as demonstrated in Ref.~\cite{Roberts2019}.

\section*{Acknowledgements}
We thank Dr.~Volker Dangendorf for providing the vacuum vessel and gas system used in these studies.
This research was supported by the Pazy Foundation, the Minerva Foundation with funding from the Federal German Ministry for Education and Research, and the Nella and Leon Benoziyo Center for High Energy Physics.
We also extend our gratitude to the Krenter-Perinot Center for High-Energy Particle Physics, the Shimon and Golde Picker–Weizmann Annual Grant, and the Sir Charles Clore Prize for their support.
A special thanks is extended to Martin Kushner Schnur for his invaluable contribution to this research.

\end{document}